\documentclass[12pt,preprint]{aastex}

\newcommand{\be}{\begin{equation}}

\newcommand{\ee}{\end{equation}}

\shorttitle{Runaway Merging in Cluster with Mass Segregation}

\shortauthors{Ardi, Baumgardt and Mineshige}

\begin{document}


\title{The influence of initial mass segregation on the runaway merging of stars}


\author{Eliani Ardi,\altaffilmark{1} Holger Baumgardt,\altaffilmark{2} and
  Shin Mineshige \altaffilmark{3}}
\altaffiltext{1}{Kyoto International University, 610-0311 Kyoto, Japan.}

\altaffiltext{2}{Argelander Institute for Astronomy, University of Bonn, D-53121 Bonn, Germany}
\altaffiltext{3}{Yukawa Institute for Theoretical Physics, Kyoto University, 606-8502 Kyoto, Japan.}

\begin{abstract}
We have investigated the effect of initial mass segregation on the runaway
merging of stars. The evolution of multi-mass, dense star clusters was 
followed by means of direct N-body simulations of up to 131.072 stars. 
All clusters started from King models with dimensionless central potentials 
of $3.0 \le W_0 \le 9.0$. Initial mass segregation was realized by varying 
the minimum mass of a certain fraction of stars whose either (1) distances 
were closest to the cluster center or (2) total energies were lowest. 
The second case is more favorable to promote the runaway merging of stars 
by creating a high-mass core of massive, low-energy stars. 

Initial mass segregation could decrease the central 
relaxation time and thus help the formation of a high-mass core.  However, 
we found that initial mass segregation does not help the runaway stellar 
merger to happen if the overall mass density profile is kept constant.
This is due to the fact that the collision rate of stars is not increased 
due to initial mass segregation. Our simulations show that initial mass 
segregation is not sufficient to allow runaway merging of stars to occur 
in clusters with central densities typical for star clusters in the Milky Way.


\end{abstract}


\section{Introduction}

The discovery of point-like, ultra-luminous X-ray (ULX) sources with 
luminosities larger than $L_X > 10^{40}~\rm{ergs~s^{-1}}$ 
by the {\it Chandra} satellite \citep{mat01,kaa01}, corresponding to a few 
hundred $M_{\odot}$ black holes (BHs) if the sources are
not beamed and accrete at the Eddington rate, 
could be a first hint for the
existence of so called intermediate-mass black holes (IMBH). IMBHs would
bridge the gap between stellar-mass BHs which form as the end-product
of normal stellar evolution and the supermassive BHs observed at the
centers of galaxies. The connection between IMBHs and ULX is also supported by 
quasi-periodic oscillations in the X-ray spectrum found in some of the sources
\citep{str03,fio04}.

Several additional arguments have also suggested the presence of IMBHs in
globular clusters (see \citet{bau05} for a review), such as (1) the extrapolation
of the $M_{BH}-M_{bulge}$ relation found for supermassive black holes in 
galactic nuclei \citep{mag98} (2) the analysis of the central velocity dispersion in the globular clusters
\object{M15} \citep{ger02} and \object{G1} \citep{geb05} (3) N-body
simulations of runaway merging of stars in young star clusters in \object{M82} \citep{por04}.

How IMBHs can form is still an open question. \citet{ebi01} proposed a
scenario in which IMBHs form through successive merging of massive stars in
dense star clusters. In a dense enough cluster, mass segregation of massive
stars is faster than their stellar evolution and the massive stars sink into
the center of the cluster by dynamical friction and form a dense inner
core. In the inner core, massive stars undergo a runaway merging process and 
a very massive star forms with a mass exceeding several 100 solar masses. 

A recent study of collisionally merged massive stars by \citet{suz07} showed that the merger products
return to an equilibrium state on a Kelvin-Helmholtz timescale
and then evolve like single homogeneous stars with corresponding mass and
abundance. The final fate of the very massive stars will depend on the assumed
mass loss rate, but IMBH formation is one possible outcome \citep{bel07}. 

Direct N-body simulations of star clusters with up to 65536 stars by
\citet{por02} showed that runaway merging can cause the formation of a massive
star with up to 0.1\% of the total cluster mass before it turns into an
IMBH. Formation of very massive stars, as progenitors of IMBHs through runaway
collisions in young star clusters has also been studied recently by
\citet{fre06a,fre06b} using Monte Carlo simulations. Utilizing a large number of
particles ($10^6$ -- $10^8$ particles), they found that runaway
  collisions could lead to formation of very massive stars with masses $\ge$
400 $M_{\odot}$.
 
Runaway merging of stars in the star cluster \object MGG-11 in the
starburst galaxy M82, whose position is consistent with a luminous X-ray
  source, has been intensively examined by \citet{por04}. They reported that
\object MGG-11 can host an IMBH if its initial dimensionless central potential
was high enough. A dimensionless central-potential $W_0 \ge 9.0$ was
required for runaway growth through collisions to form an IMBH. Unfortunately,
such a high dimensionless central potential leads to a central density $\rho_c
\ge 10^6 M_\odot/{\rm pc}^3$ which is rarely seen in present-day star
clusters, implying that the formation of IMBHs in star clusters is a very rare
event.

One possible way which would allow runaway collisions to occur in clusters
with lower central density is the assumption of initial mass
segregation. Initial mass segregation, which allows massive stars to start
their life in the cluster center, might be a way to lower the density
requirement for the onset of runaway collisions. The
tendency for massive stars to form preferentially near the cluster center is
expected as a result of star formation feedback in dense gas clouds
\citep{mur96} and from competitive gas accretion onto protostars and mutual
mergers between them \citep{bon02}. Observational evidence for initial
mass segregation in globular clusters as well as in open clusters has also
been reported \citep{bon98,deg04}.

Dynamical evolution of young dense star clusters with initial mass segregation
until the onset of core-collapse stage has been studied by \citet{gur04} by
using Monte Carlo simulations. Besides decreasing the core collapse time, they
found that initial mass segregation applied in clusters with $N = 1.25 \times 10^6$ stars which
followed a Plummer density profile initially, results  
in a total mass of the collapsed core of about 0.2 \% of the total cluster mass.

Motivated by results of \citet{por04}, that without initial mass segregation,
the dense star cluster MGG-11 could experience runaway merging only if the
central density was higher than $10^6 M_\odot/{\rm pc}^3$, in the present
study we want to explore whether or not initial mass segregation could lower
the density required for runaway collisions in \object MGG-11 like
clusters.  For this purpose, we perform $N$-body
simulations of \object MGG-11 like clusters starting from different initial
conditions which are described in detail in the next section. Results and
analysis of our simulations are shown in section~3 while the discussion and
conclusions are presented in section~4.

\section{Details of numerical simulations}\label{sec:ns}

We have conducted a number of $N$-body simulations, using the collisional
$N$-body code NBODY4 (Aarseth 1999) on the GRAPE-6 special purpose computers
provided by ADC - CfCA NAO Japan, to follow the evolution of multi-mass star
clusters. All simulations are run for a time span of 3 Myrs by which time 
we assume that the
runaway stars are turned into BHs and stop the simulations.

Our clusters contain 131.072 stars initially, distributed according to a
Salpeter IMF with minimum mass and maximum mass equal to 1.0 
$M_{\odot}$ and 100 $M_{\odot}$ respectively, which is chosen to fit the
\citet{mcc03} observations for MGG-11. Stellar evolution is modeled   
according to \citet{hur00}. Since we only follow the first 3 Myrs of cluster evolution, 
stellar evolution is important only for the most massive stars. Two stars are assumed to 
'collide' if the distance between them becomes smaller than the sum of 
their radii. We assume that the total mass of both stars ends in the merger 
product and do not follow the stellar evolution of the runaway stars. We 
examine the evolution of \citet{kin66} models with central 
concentration $3.0 \le W_0 \le 9.0$. The initial half-mass 
radius and total cluster mass are chosen similar to what \citet{por04} chose 
to fit the observed parameters of MGG-11, namely $r_{h} = 1.3~\rm{pc}$ and $M
= 3.5 \times 10^5 M_{\odot}$. Details of the simulated
clusters without initial mass segregation are presented in Table~1.

In order to examine the effect of initial mass segregation, we study   
two scenarios. In the first scenario, we vary the minimum mass 
$m_{min}$ within the lagrangian radius containing 5\% of the total 
cluster mass ($R_{005}$). Increasing the minimum mass $m_{min}$ within 
$R_{005}$ (from 1 $M_\odot$ to $m_{min} > 1 M_{\odot}$ for clusters with 
initial mass segregation), while keeping the total cluster mass and energy 
constant, will consequently decrease the number of stars within this 
sphere. This scenario allows massive stars to start their life in the 
cluster center. It is proposed to meet observations which show that massive 
stars are preferentially formed near the cluster center \citep{bon98,deg04}. 
Details of runs where mass segregation 
is introduced inside a certain radius are given in Table~2.

In the second scenario, we choose a certain fraction of stars 
(whose total mass is 5 \% -- 20 \% of total mass of the cluster) with 
the lowest total energy 
and then vary the minimum mass of them, while keeping the total cluster 
mass and energy constant. The number of stars is again lower than in a normal 
cluster. Compared to the first scenario, the second scenario brings 
massive stars even closer to the center since stars located in the center at
time t=0 could still have high energies and spent most of their life 
outside the center. Hence support for runaway collisions should be stronger 
in the second scenario.

We also vary the half-mass radius of the clusters to see the effect of 
different central densities.
Table~3 reports details for clusters with initial mass segregation,  using
  the second scenario.

\section{Results and Analysis}\label{sec:sim}

\subsection{Clusters without initial mass segregation}

We run five cluster models without initial mass segregation as shown 
in Table~1. Each cluster contains 131.072 stars, but has different $W_0$. 
Four of them are set to have the same half-mass radius, which is 1.3 pc, 
to mimic MGG-11. In addition, we also examine a $W_0$ = 7.0 cluster 
with a smaller half-mass radius of $r_h$ = 0.5 pc. The central density of 
each cluster refers 
to the density within the core radius of the cluster, which is determined
with the method of \citet{cas85}.
For clusters with the same $r_h$, the central density is higher 
for clusters with higher dimensionless central potential $W_0$. 

We also calculate the central relaxation time of each cluster to study the 
influence of this parameter on the occurrence of runaway merging. 
The central relaxation time $T_{rel,c}$ is defined as \citep{spi87}:
\begin{equation}
T_{\rm rel,c}= \frac{\sigma^3_{3D}}{4.88 \pi G^2 \ln(0.11 N)n \langle m
  \rangle^2} ,
\end{equation}
where $\sigma_{3D}$, $n$ and $\langle m\rangle$ are the three-dimensional
velocity dispersion, number density and average stellar mass at the cluster
core. Here the cluster core refers to the region inside the core radius
$\rm r_{core}$. 

Our simulations of MGG-11 like clusters (with $r_h$ = 1.3 pc) show (see
Table~1)  that 
only the star cluster with the highest dimensionless central potential ($W_0$ = 9.0, 
corresponding to a central density of 3.24 $\times 10^6~M_{\odot}/{\rm{pc^3}}$),
experiences runaway merging. This result is in a good agreement with 
the one found by \citet{por04}. Our result again proves that high central 
density is required to allow runaway merging to occur. Collisions among
massive stars also occur in the lower density cluster but none of them 
experiences subsequent collisions leading to a super-massive star.

Fig.~1 depicts the evolution of lagrangian radii containing 1\% -- 20 \%
  of total mass of cluster models~1~--~3. Core radii ($\rm r_{core}$), which
  are marked by bold lines, are calculated according to \citet{cas85}. 
The inner shells of the $W_0= 9.0$ cluster (model~1) suffer strong contractions 
due to the high central density. Core collapse happens in this cluster at 
$t \approx$ 0.6 Myrs. The core collapse supports the runaway merging to
happen since runaway merging sets in at $t = $ 0.54 Myrs, about the same
time when core collapse happens (see the 10th column of Table~1.). Inner shells of $W_0$ = 7.0 cluster 
(model~2), on the other hand, contract very slowly. Even until 3 Myrs, the contraction is not 
strong enough to produce core collapse. Consequently, no runaway merging 
occurs in this cluster. Evolution of inner shells of $W_0$ = 7.0 cluster 
however looks different when we decrease $r_h$ to 0.5 pc (model~3). 
Mild contraction brings the cluster to collapse. Core collapse occurs at 
$t \approx $ 2.6 Myrs. At the same time, the first collision
leading to runaway merging happens ($t = $ 2.55 Myrs, see the 10th column of
  Table~1). 
Although the runaway merging started later than in the $W_0 =$ 9.0 
cluster, three collisions are enough to form a super-massive star  
with few hundred $M_\odot$ (see columns 9 and 11 of Table~1). 

The two clusters which experience runaway merging (models~1 and ~3), 
have very high central densities. The $W_0$ = 7.0 model has 
$\rho_c$ = 2.95 $\times 10^6~M_{\odot}/\rm{pc^3}$ while 
the $W_0$ = 9.0 model has $\rho_c$ = 3.24 $\times 10^6~M_{\odot}/\rm{pc^3}$. 
Runaway merging does not occur in clusters whose central densities are lower 
than $10^6~M_{\odot}/\rm{pc^3}$. 
Therefore the critical density which allows clusters without initial mass 
segregation to experience runaway stellar merging should be larger than 
$10^6 M_{\odot}/\rm{pc^3}$. This limit holds for globular-cluster-size 
objects with masses of $10^5 M_{\odot}$. 

Since in our runs the central density is varied, we find that
the central relaxation time (see column 6 of Table 1) mainly depends on the
number density of stars in the center, where $T_{rel,c} \propto n^{-1}$ (see
eq. 1). The central relaxation time is hardly affected by the change of
velocity dispersion $\sigma$ and average mass $\langle m \rangle$ (on average
$\sigma \approx$ 27.9 km/s and $\langle m \rangle \approx$ 2.64 $M_{\odot}$.) A high number density of stars in the cluster center seems to be required
to support runaway stellar merger in a cluster without initial mass
segregation.

Our result, that runaway merging does not occur in clusters with too low central
  density, is in good agreement with the one found by
  \citet{fre06a,fre06b}. Fig.~1 of \citet{fre06b} (which is essentially the
  same as Fig.~1 of \citet{fre06a}) shows that a cluster with mass 3 $\times$
  $10^5 M_{\odot}$ and dimensionless central potential $W_0 = 8.0$ experiences
  runaway collisions if its N-body length unit $(R_{NB})$ $\le$ 2 pc. This
  value corresponds to an initial half mass radius $R_h \le$  1.74 pc (see 
  sec.~2.1 of their paper where they show that $R_{NB} \simeq 1.15~R_h$ for $W_0 =
  8.0$). This value is not too far from the critical value we find, since our
  simulations show that a $W_0 = 7.0$ cluster without initial mass segregation experiences
  runaway collisions when its initial half-mass radius is somewhere between 1.3
  pc and 0.5 pc (see models~2 and 3 in Table~1), while a $W_0 = 9.0$ cluster
  with initial half-mass radius 1.3 pc experiences runaway collisions.


\subsection{Clusters with initial mass segregation}

In models 6 -- 8, we introduce initial mass segregation by replacing stars
within the 
5 \% lagrangian radius ($R_{005}$) with massive stars whose masses are higher 
or equal than the mass $m_{min}$ written in the 6th column of Table~2.
Replacing is done by randomly selecting new positions and velocities
for the massive stars from the positions and velocities of innermost stars. The
number of massive stars is chosen such that the overall mass density
profile remains constant.

As we keep the mass within the $R_{005}$ lagrangian radius constant, 
introducing initial mass segregation by increasing $m_{min}$ means to 
increase the average mass $\langle m\rangle$ of stars and lower the 
total number of stars (see the 3rd column of Table~2). The increase of
$\langle m\rangle$ in this region consequently decreases the central 
relaxation time $T_{rel,c}$. The central relaxation time of these clusters
should be lower than the one of a $W_0$ = 7.0 cluster without initial mass
segregation (see column 6 of model~2 in Table~1). As the central parts of these clusters
relax faster, the clusters may evolve faster and core collapse could
happen earlier. One may therefore expect that runaway merging should
now occur at lower central densities.

The top part of Fig.~2 shows that model~6 (with $m_{min} =
  30~M_{\odot}$) does not experience core collapse before 3 Myrs. 
Even increasing $m_{min}$ up to
$90~M_{\odot}$, as in model~8, does not
lead the cluster to experience core collapse before 3 Myrs either
(see bottom part of Fig.~2). Our simulations also show that no
runaway merger occurs in these clusters.
The reason why runaway merging does not happen is that
massive stars, which start their life in the region within $R_{005}$ do not 
constantly stay there. Some of these massive stars, whose initial
velocities are high enough, leave this region. Since the cluster is initially
mass segregated, this outward movement of massive stars is not balanced by a sufficiently
large number of massive stars moving inward, hence the average mass of stars 
decreases in the center. We note however that, since our clusters are started in
virial equilibrium, the expansion of the
high-mass stars is balanced by a corresponding number of low-mass stars moving further in, so that
the central density remains constant. 

The depletion of massive stars from the initial $R_{005}$ is shown for model~6 in 
Fig.~3. This figure depicts the evolution of lagrangian radii of 
massive stars whose masses are higher than 30 $M_{\odot}$ and that started 
their life inside $R_{005}$. Total mass fraction of these massive stars is
indicated by $M_{005}$. The lagrangian radii containing between 10 \% up to 
100 \% of these stars are presented. 
The upper figure shows the change of lagrangian radii within the first 
0.05 Myrs. We can see that within a few core crossing times ($
  t_{cross} \approx 8
  \times 10^3 \rm{yrs}$ ) some of these 
massive stars leave the initial $R_{005}$. At t=0.05 Myrs, total mass of massive stars 
which still reside inside this region is only 60 \% of the initial mass 
$M_{005}$. Bottom figure shows that 
up to t=3 Myrs, this region contains only about 30 \% of total mass 
of these massive stars.

Increasing
the minimum mass of stars whose distances are closest to
the cluster center does not succeed to produce high-mass
cores. In order to keep massive stars in the cluster core, we used a second 
scenario where initial mass
segregation is realized by varying the minimum mass of a certain fraction
of stars whose total energies are lowest. Since the second scenario is more
favorable to create a high-mass core of massive, low-energy stars, we will base
our results on this scenario.

In the second scenario, initial mass segregation was
introduced by replacing
stars which have the lowest total energy, up to 5 \% -- 20 \% of the total 
mass of the cluster (models 9 -- 15, see $M_{IMS}$ in column~5 of Table~3)
with massive
stars whose masses are higher than $m_{min}$. The coordinates and 
velocities of massive stars are randomly chosen from the stars with the lowest total energy
and their total number is again adjusted such to keep the overall
mass density profile constant and the cluster in virial equilibrium.

In order to show that clusters are in virial equilibrium, Figs. 4 and 5 depict the evolution 
of lagrangian radii of all stars and those of massive (M $\ge$ 30 $M_{\odot}$) and less massive 
stars (M $< 30~M_{\odot}$) of cluster model~11. As can be seen lagrangian radii of 
massive as well as less massive stars are nearly constant within the  
first few crossing times. This shows that the cluster is in a stable equilibrium
condition after mass segregation was introduced.

The central density and central relaxation time are measured for the
region inside the cluster core. Since massive stars
are not strongly concentrated toward the cluster center, the mean mass 
of stars within the cluster core is not very high (4.77 $M_{\odot}$ -- 11.82 $M_{\odot}$). 
Therefore the central relaxation time of clusters with $r_h$ = 1.3 pc (6.55 $\rm{Myrs}$ $\le T_{rel,c} \le$ 8.69 $\rm{Myrs}$) 
is not as low as that of the clusters in Table~2 
(3.54 $\rm{Myrs}$ $\le T_{rel,c} \le$ 3.84 $\rm{Myrs}$).
One may expect that the central relaxation time should be short enough that 
massive, low-energy stars spiral into the cluster core and create 
a high-mass core. Once in the cluster core, these massive stars could 
collide with each other and promote runaway merging.   

Nevertheless, our simulations do not show runaway merging (see models~9 -- 12
of Table~3). Reducing the half-mass radius $r_h$
from 1.3 pc to 0.5 pc in order to increase the central density 
(models~13 -- 15, see column 4 of Table~3) does not
help runaway merger to occur either.  

Model~15 actually has the same initial central density and half-mass radius as
model~3. Initial mass segregation is not introduced in model~3, but the
cluster experiences runaway merging through three collisions (see
Table~1). Fig.~6 depicts the evolution of lagrangian radii of inner shells of
these two models. Both clusters experience contractions of their
cores. While the contraction of model~3 is sufficiently strong to
let core-collapse occur at $t=2.6$ Myrs, the core of model~15 does not 
collapse until 3 Myrs and no runaway merging occurs.

By using the Monte Carlo method, \citet{gur04} studied core-collapse of star
clusters with initial mass segregation. A direct comparison of their results
with ours is again difficult due to differences in the adopted initial mass
spectrum, density profile, number of particles (up to $N = 10^7$ for
\citet{gur04}) and the method used in introducing initial mass
segregation. \citet{gur04} note in the caption of Fig.~13 that stellar
evolution can reverse core collapse. This agrees at least qualitatively with what
we see in our runs, since for example Fig.~6 shows that, despite of similar
size and density profile, model 3 goes into core collapse earlier than model
15. This could be due to the fact that core collapse in model 15, whose core
contains many high-mass stars due to initial mass segregration, is delayed by
the stronger mass loss from the core due to stellar evolution.

Besides the effect of stellar evolution, the difference of the evolution of
models~3 and~15 may due to the difference of their collision rates. We examine the collision rate of
these models by calculating the the collision rate $N_{Coll}$ using equation (8-122) of \citet{bin87}
.

\begin{equation}
N_{Coll\star} = 4\sqrt{\pi} n \sigma {(2~R_{\star})}^2 + 4 \sqrt{\pi} G M_{\star} n (2~R_{\star})/\sigma.
\end{equation}

Here $N_{Coll\star}$ is the average number of collisions that a star suffers 
per unit time, $n$ indicates the number density of stars,
$\sigma$ is the velocity dispersion of stars, 
$R_{\star}$ and $M_{\star}$ denote radius and mass of colliding stars, 
and $G$ is the gravitational constant. The first term is derived from the 
kinetic theory for inelastic encounters and the second term represents the
enhancement in the collision rate by the gravitational attraction of the two
colliding stars.

Let us consider the region inside the core radius $\rm r_{core}$. The average number of collisions per unit time $N_{Coll}$ 
is obtained by multiplying $N_{Coll\star}$ with the number of stars inside 
the core radius $N_{\star core}$. Therefore 
\begin{equation}
N_{Coll} = N_{Coll\star}~N_{\star core}.
\end{equation}
The number density
of stars inside the core radius $n$ can be written as
\begin{equation}
n = N_{\star core} / V_{core}. 
\end{equation} 
where $V_{core} = \frac{4\pi}{3} \rm r_{core}^3$.
Thus
\begin{equation}
N_{Coll} = N_{Coll \star}~n~V_{core}.
\end{equation}
Substituting $N_{Coll\star}$ with the expression written in eq.~2, we see that
\begin{equation}
N_{Coll} \propto n^2.
\end{equation}

We use the theoretical prediction of the collision rate (eqs.~2 and~5) to 
follow the growth in the number of collisions per unit time in models~3 and 15. 
$N_{Coll\star}$ is calculated by considering the mass and radius of 
each star and then summing up over all stars within the region inside
the core to obtain $N_{Coll}$. Core parameters and collision rates
are calculated each time $N$-body data was stored and are then summed up
over all times.

The theoretical estimates are compared with the collision rate we
find in our simulations in Fig.~7. Both theoretical and simulation results 
(see column 7 of  Table~1 and column 9 of Table~3)
show that the collision rate of model~3, which
experiences runaway merging, is higher than the one in model~15.
The theoretical prediction of the collision rate overestimates the simulation 
results by a factor $\approx$ 2. This may be due to assumptions 
(i.e. mass and radius of colliding stars are the same) and idealizations 
(i.e. distribution function of velocity is Maxwellian) used in the 
derivation of eq.~2, while in the
simulations we use a mass spectrum and stellar radii according to
a certain mass-radius relation.

\section{Discussion and Conclusions}\label{sec:discon}

We have followed the evolution of multi-mass, dense star clusters with 
dimensionless central potentials of 3.0 $\le W_0 \le$ 9.0. Our simulations 
show a good agreement with the results of \citet{por04} that in MGG-11 type clusters
without initial mass segregation, dimensionless central potentials $W_0 \ge 9.0$
corresponding to  
central number densities larger than $10^6/\rm{pc}^3$ are required 
for runaway mergers to occur. Examining clusters with lower dimensionless central potential,
$W_0 \le 7.0$, confirm this limit for runaway mergering, as shown in Fig.~8. 

Initial mass segregation increases the average mass of stars within 
the cluster center and thus decreases the central relaxation time. It 
also allows to form a high-mass core. However, as long as the mass density
profile is kept constant, we find that 
initial mass 
segregation does not support runaway stellar merging to happen since
the collision rate is decreased.

In spite of the differences in adopted IMF, number of particles, treatment of
stellar evolution and stellar collisions, our results are in line with
\citet{fre06a,fre06b} (see sec.~3.1) and \citet{gur04} (see sec.~3.2).

The data of Milky Way globular cluster given by \citet{har96} provide
the central luminosity density (in $L_{\odot}/pc^3$) which can be converted
into a central mass density by assuming a mass-to-light ratio $M/L = 1$. 
Doing this, we find that about 67 \% of Milky Way
globular clusters have central densities $10^2~M_{\odot}/\rm{pc^3} < \rho_c <
10^5~M_{\odot}/\rm{pc^3}$. Only 4 \% have central densities exceeding $\rho_c
= 4.3 \times 10^5~M_{\odot}/\rm{pc^3}$, while none has a central density
larger than $\rho_c = 10^6~M_{\odot}/\rm{pc^3}$, as depicted on Fig.~9. 
Studies of the evolution of clusters containing IMBHs by \citet{bau04a,bau04b} have 
shown that clusters with IMBHs expand due to energy generation in their cusp. 
\citet{bau04b} found that the 
cluster expansion can be strong enough that very concentrated clusters can end up 
among the least dense clusters. However, Milky Way globular clusters have 
half-mass radii very similar to the radii of clusters which form today, like 
galactic open clusters or super-star clusters in interacting galaxies 
(see i.e. \citet{sch06,tra07}), which speaks against strong expansion. 
If the current densities are representative 
of the densities with which the clusters formed, then runaway merging would not have 
happened in any of these clusters. In addition, the data of young star clusters in the LMC given by
\citet{mac03} (see Table 6 of their paper) also shows that nearly all LMC 
clusters, including very young ones, have central densities far below the
critical value needed for runaway merging. Other possibilities
of forming IMBHs like the merging of many stellar mass black holes 
\citep{mil02} also need extreme initial conditions like very massive 
clusters \citep{gul04,ras06}.

Hence it seems likely that most star clusters did not have 
sufficient high central densities to form IMBHs.
This indicates that the formation of IMBHs in star 
clusters must have been a rare event.   

\acknowledgments

We thank for Douglas Heggie for valuable discussions. This work was
supported in part by the Grants-in-Aid of the Ministry of Education, Culture,
Sports, Science and Technology, Japan,~(14079205; EA,SM). Numerical computations
were carried out on GRAPE system at Center for Computational Astrophysics of
National Astronomical Observatory of Japan. Data analysis were in part carried
out in YITP, Kyoto University.

\begin{figure}
\includegraphics[scale=0.8]{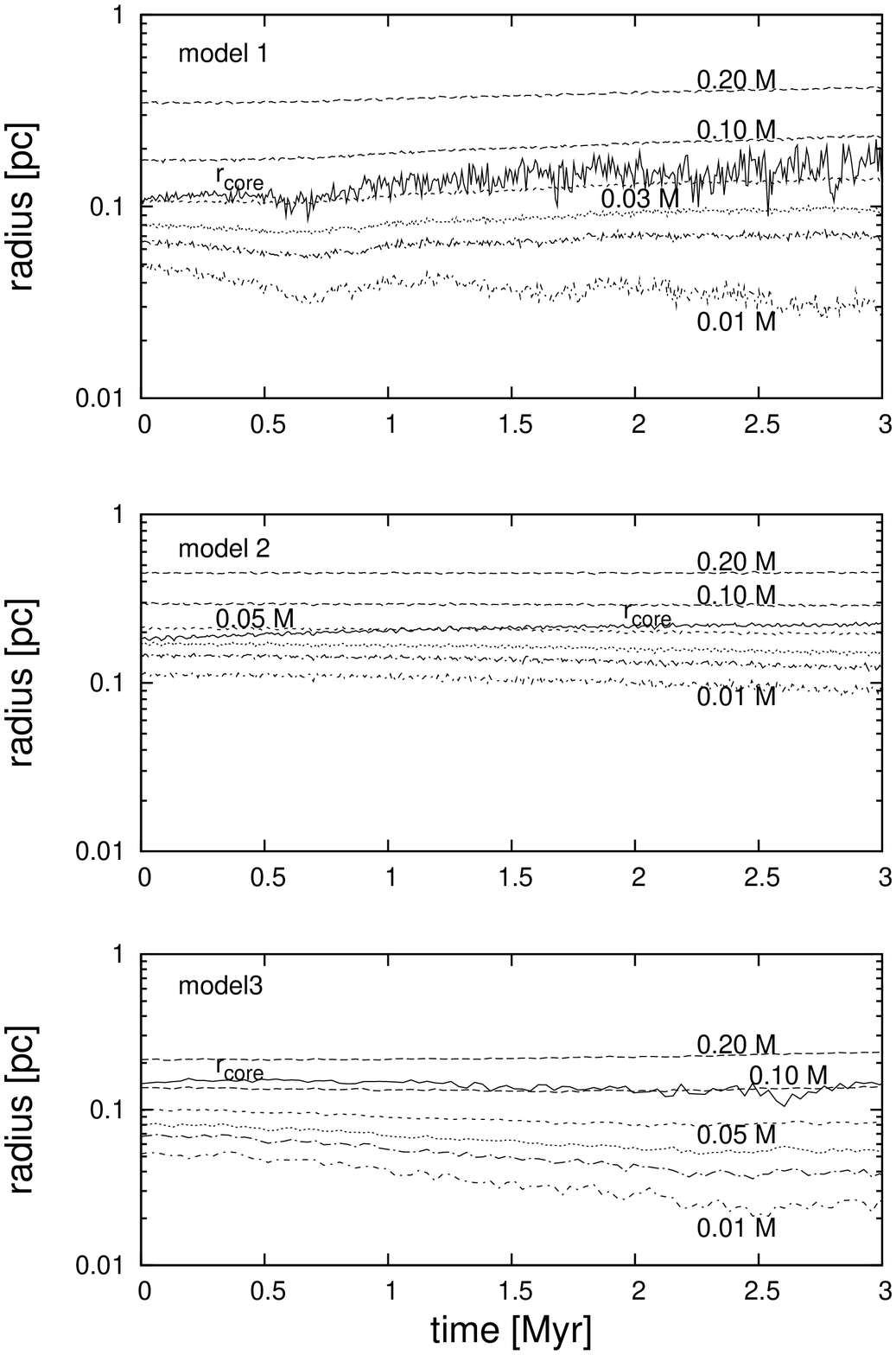}
\caption{Evolution of lagrangian radii of inner shells containing 1\%, 2\%,
  3\%, 5\%, 10\% and 20\% of the total cluster mass of models~1 --~3. Core
  radii $\rm{r_{core}}$ are marked by bold lines. Model~1 has short 
enough central relaxation time that core collapse and subsequent runaway 
merging of stars happen within a few Myrs. Model~2 has a higher $T_{rel,c}$
(see column 6 of Table~1) which prevents its core to collapse. Compared to model~1, 
model~3 has similar value of $\rho_c$ which allows mild contractions 
to bring the core to collapse before 3 Myrs. 
  \label{fig1}} 
\end{figure}

\begin{figure}
\includegraphics[scale=0.8]{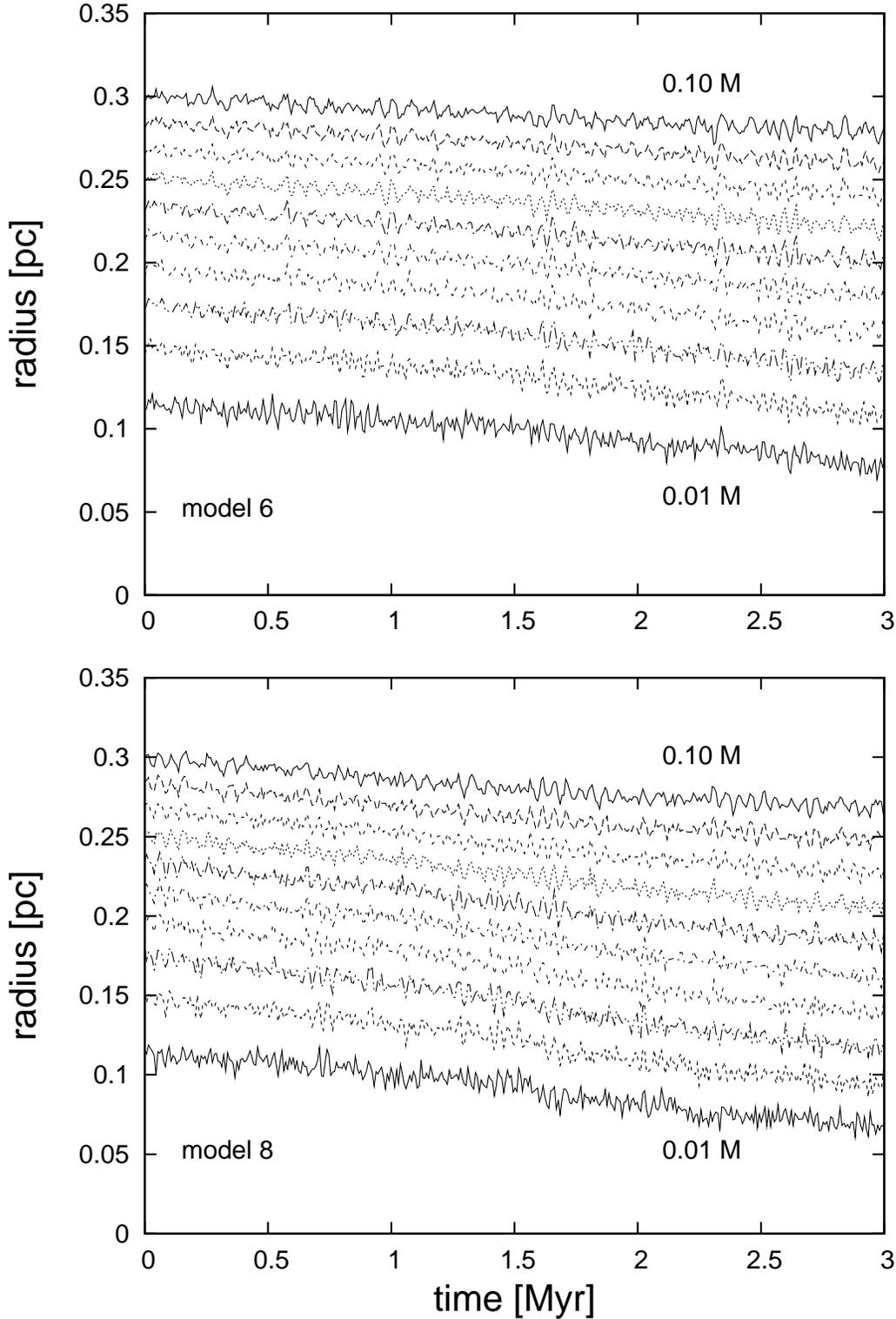}
\caption{Evolution of lagrangian radii of inner shells containing 1\% -- 10\% 
of the total cluster mass of model~6 (top) and model~8 (bottom). Filling the
region inside the 5 \% lagrangian radii $R_{005}$ with stars more massive than
30 $M_{\odot}$ (model~6) does not support the core to collapse. Even
increasing the minimum mass of stars in this region to 90 $M_{\odot}$
(model~8) does not help core collapse to happen. The reason is that a large
fraction of massive stars, which start their life inside the $R_{005}$, move
out of this region on a crossing time-scale (see Fig.~3).
  \label{fig2}} 
\end{figure}

\begin{figure}
\includegraphics[scale=0.8]{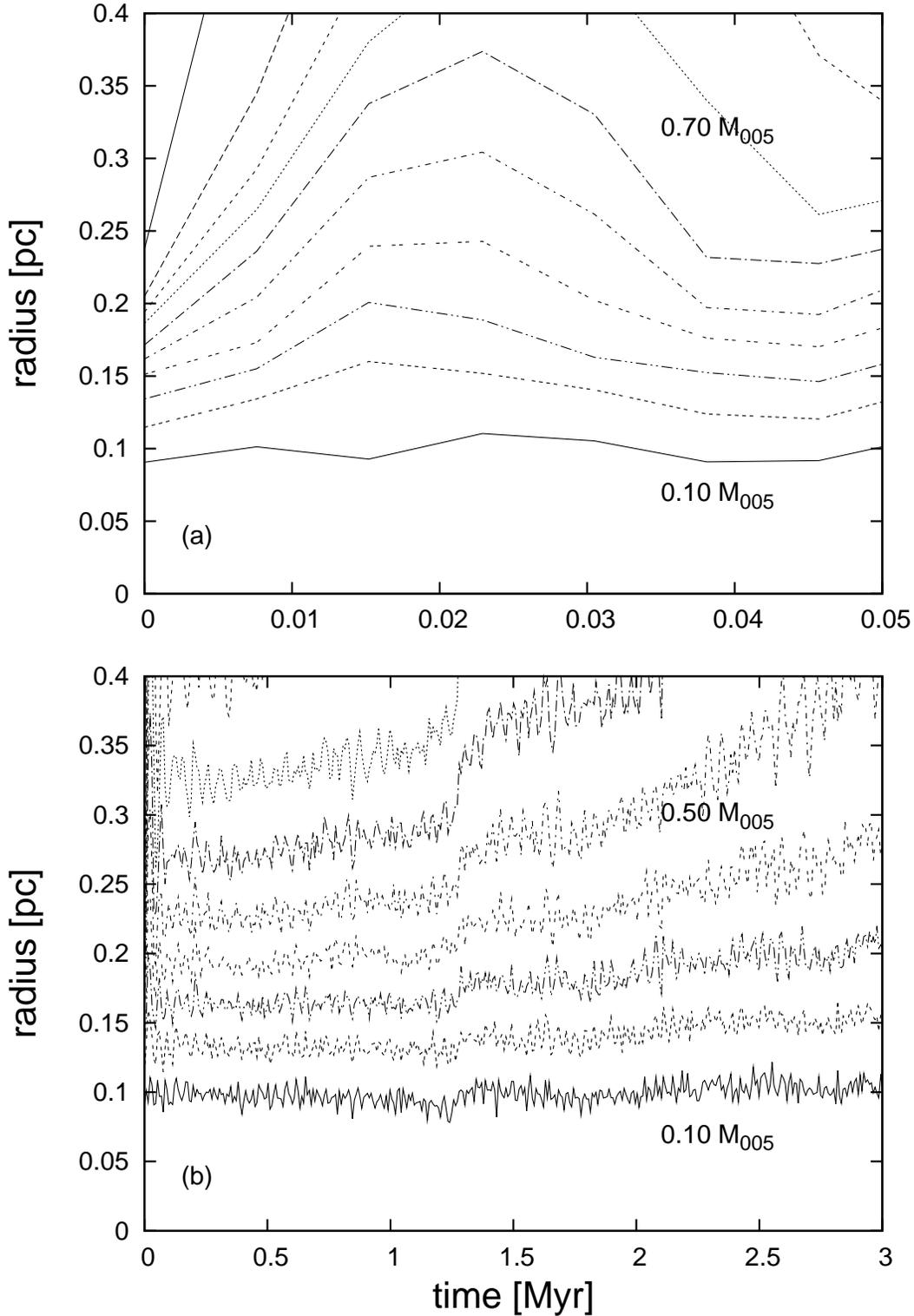}
\caption{Evolution of lagrangian radii of massive stars ($m_{min} = 30
  M_{\odot}$) which start their life inside the 5 \% lagrangian radius of the
  cluster model~6 (a) up to the first 0.05 Myrs, (b) until 3 Myrs. Total mass
  fraction of these massive stars is indicated by $M_{005}$. Within a crossing
  time-scale, some massive stars leave the region within the initial 5\%
    lagrangian radius, which is 0.24 pc in this model,
  due to their high initial velocities. The escape of the
massive stars is balanced by low-mass stars moving in from larger radii
(which is not shown in these figures).
  \label{fig3}} 
\end{figure}

\begin{figure}
\plotone{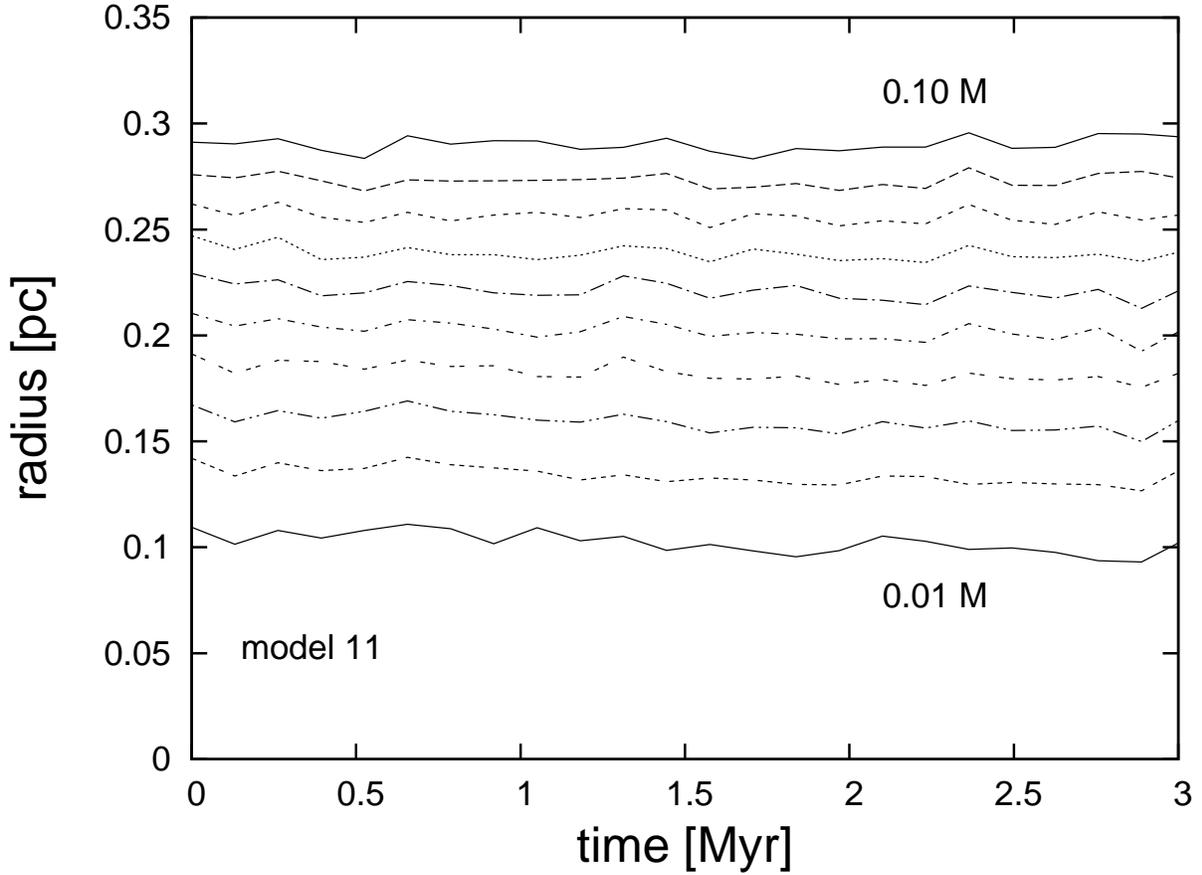}
\caption{Evolution of lagrangian radii of inner shells containing 1\% -- 10\% 
of the total cluster mass of model~11. Replacing 10\% of the lowest total
energy stars with stars more massive than 30
  $M_{\odot}$ does not support the core to collapse. This happens because the
  high-mass core did not be form until 3 Myrs.
  \label{fig4}} 
\end{figure}

\begin{figure}
\includegraphics[scale=0.8]{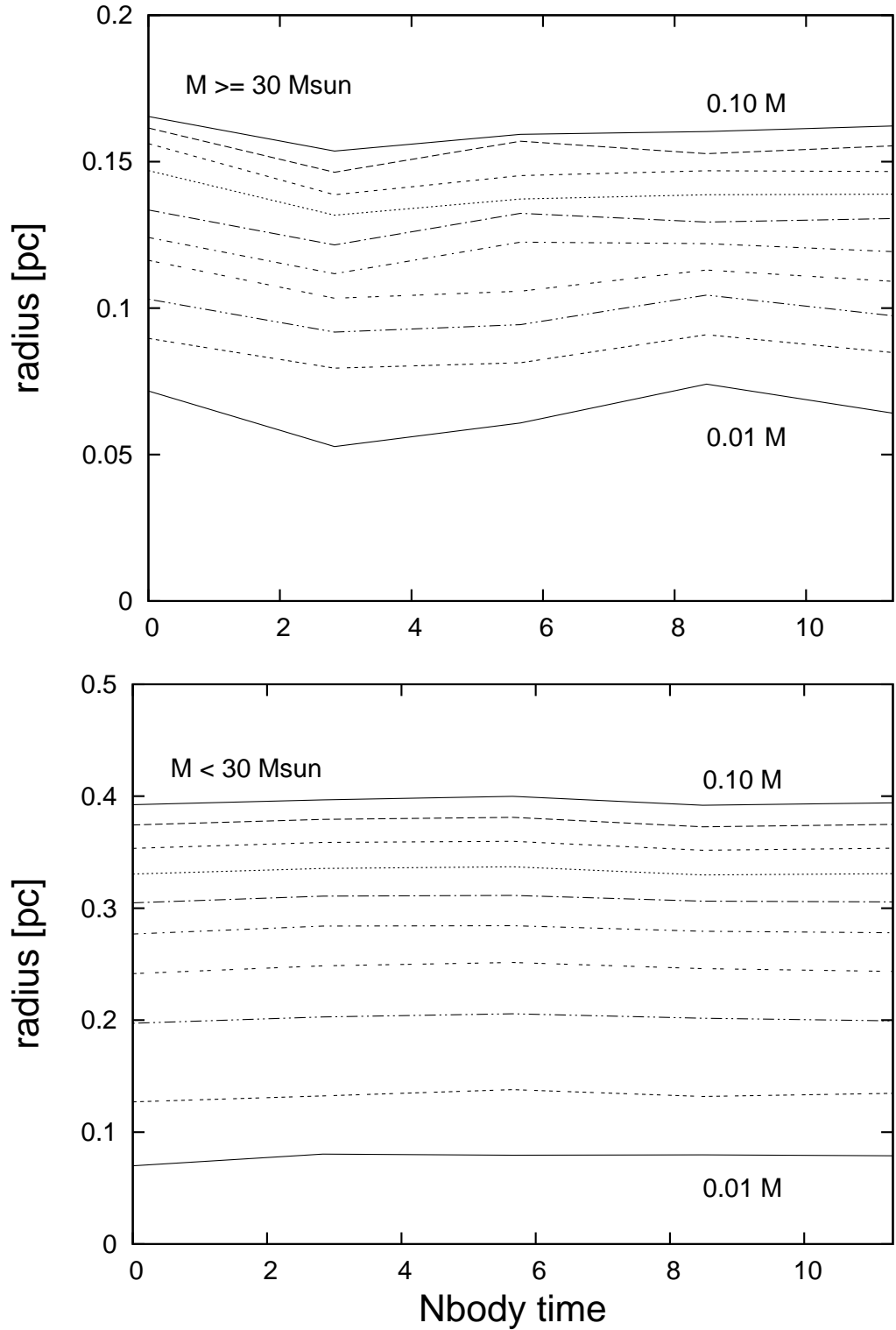}
\caption{Evolution of lagrangian radii of massive stars whose masses are at
  least 30 $M_{\odot}$ (top) and low-mass stars whose masses are less
  than 30 $M_{\odot}$ (bottom) up to 10 Nbody unit. These figures depict stable
  evolution of shells containing 1\% -- 10\% 
of the total mass of these stars in the cluster model~11. 
  \label{fig5}} 
\end{figure}

\begin{figure}
\includegraphics[scale=0.8]{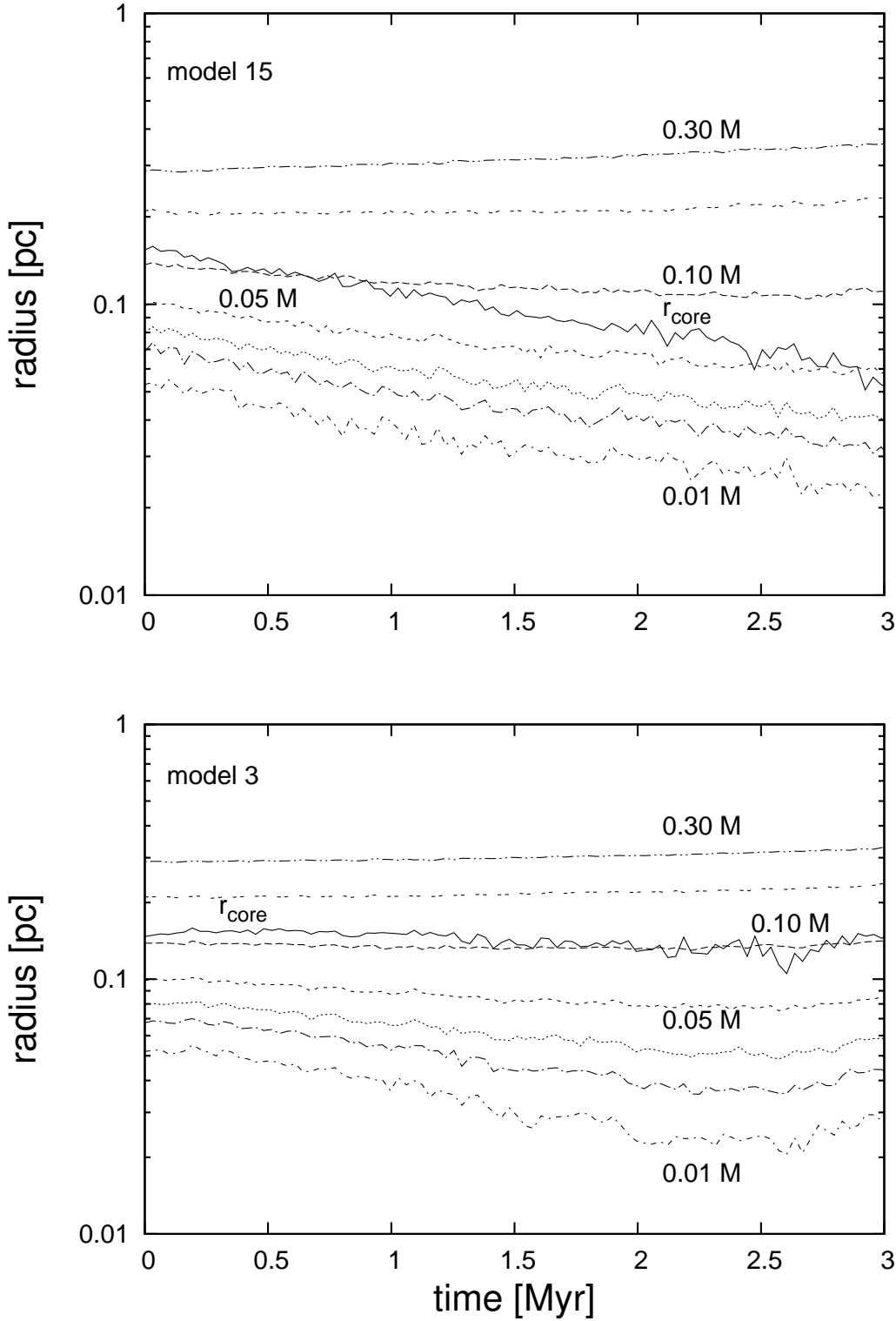}
\caption{Evolution of lagrangian radii containing 1\% -- 30\% of total
  mass of model~15 and model~3. Initial
  mass segregation is applied in model~15 by replacing 20 \% of stars with
  the lowest total energy by massive stars with $m_{min}$ = 30
  $M_{\odot}$. The inner shells experience contraction but no core collapse
  until 3 Myr. Therefore runaway merger does not occur in this cluster. The cluster model~3 has same initial density profile and
  same half-mass radius as model~15, but no initial mass
  segregation. However, mild contraction in the inner shells of model~3 is
  enough to let runaway mergings occur. This may happen since the number of
  collisions inside the inner shells of model~3 is higher than the one in
  model~15 (see Fig.~7). 
  \label{fig6}} 
\end{figure}

\begin{figure}
\plotone{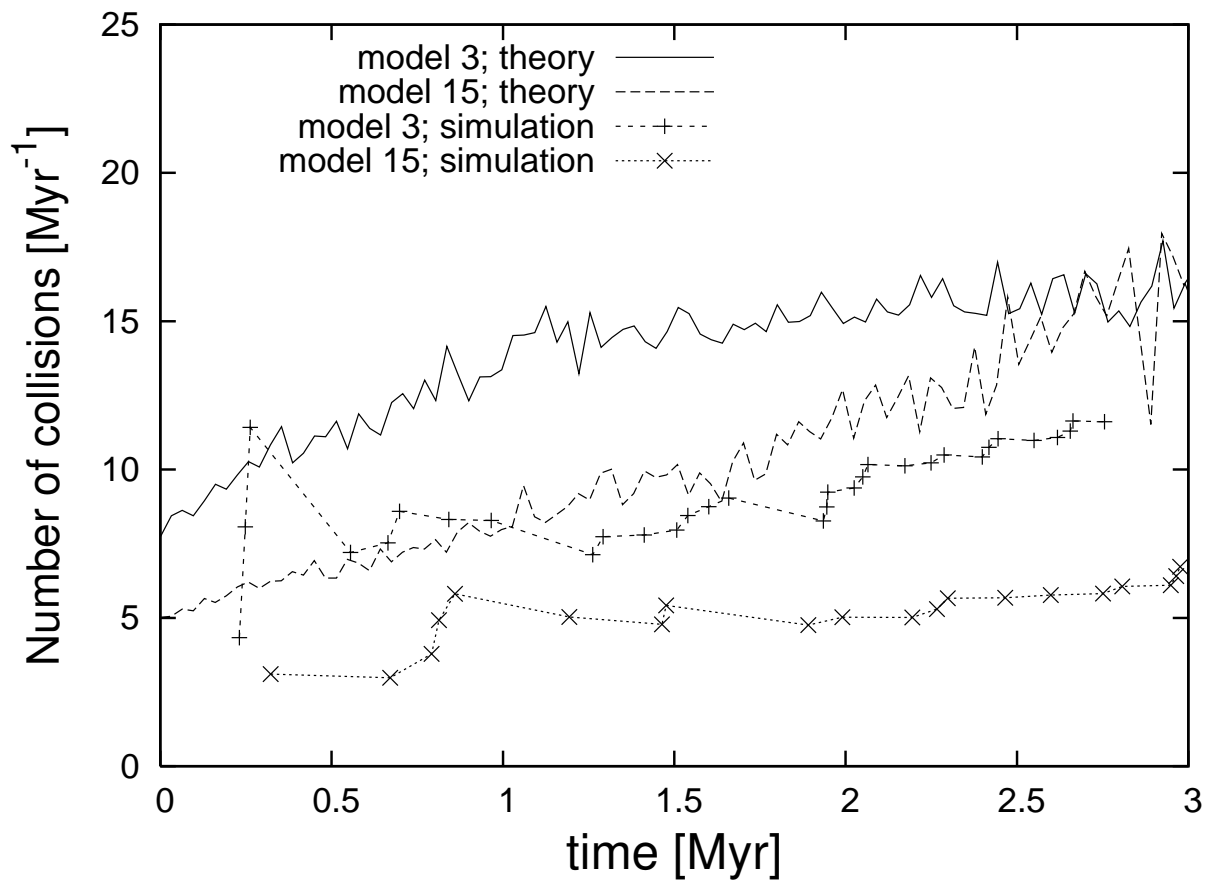}
\caption{Collision rate inside inner shells of cluster models~3 and~15
  obtained from simulations, compared to the theoretical prediction of
  collision rate based on inelastic encounters. The collision rate of the
  model without mass segregation (model~3) is higher than the model with
  initial mass segregation (model~15) because there are more stars inside the
  cluster core. Therefore the possibility for a runaway merger to occur
  is also higher. 
  \label{fig7}} 
\end{figure}

\begin{figure}
\plotone{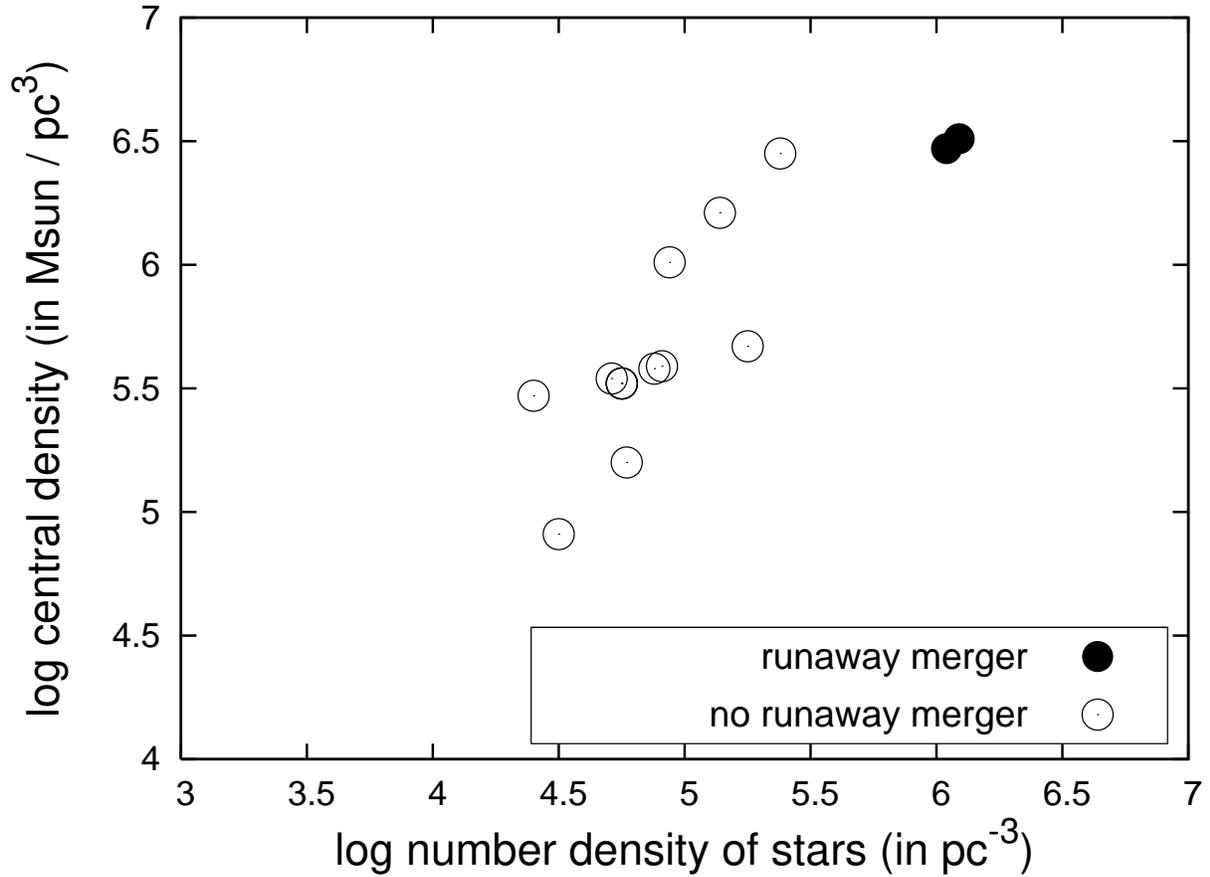}
\caption{Plot of log central density vs. the log number density of stars 
for all calculated models. In order for runaway mergers to occur, a number
density of stars larger than $10^6/\rm{pc}^3$ in the core is necessary.
  \label{fig8}} 
\end{figure}

\begin{figure}
\plotone{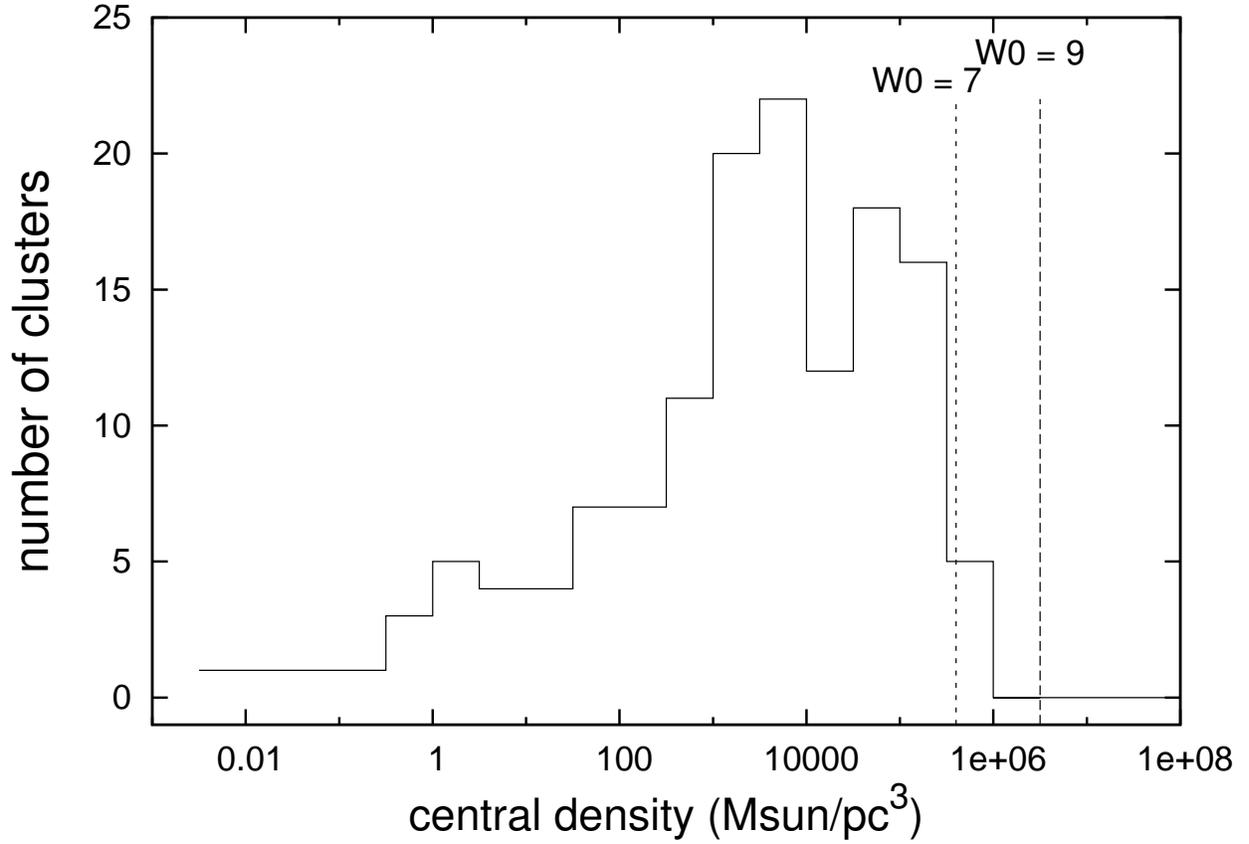}
\caption{Distribution of central densities of Milky Way globular clusters. The 
dashed lines mark the central densities of clusters in which runaway merging 
occured (models~1 and 3 of Table~1). Most galactic globular clusters have 
central densities far below this limit, meaning that runaway merging of stars 
was unlikely to have occured in them.
  \label{fig9}} 
\end{figure}

\onecolumn
\begin{table}
\begin{center}
\caption{Properties of simulated clusters without initial mass segregation \label{tab1}}
\begin{tabular}{cccccccccccc}
\tableline\tableline
 ${}_1$ & ${}_2$ & ${}_3$ & ${}_4$  & ${}_5$  & ${}_6$  & ${}_7$  & ${}_8$
 & ${}_9$ & ${}_{10}$   & ${}_{11}$ & ${}_{12}$ \\
 $Model$ & $W_0$ & $N_{Star}$ & $r_h$ & $log~\rho_c$ &
$T_{\rm rel,c}$ & $Col$ & $\langle T_{\rm col}\rangle$ & $Col_{\rm rm}$ & $
 T_{\rm rm}$ & $M_{RS}$ & $RM$ \\
       &    &         & $(pc)$  & $(M_{\odot}/pc^3)$ &
$(Myr)$         &     & $(Myr)$   &     &  $(Myr)$     &  $(M_{\odot})$ & $(Y/N)$ 
 \\ 
\tableline
 1 & 9.0 & 131072 & 1.3 & 6.51 & 1.16 & 104 & 0.03  & 96 & 0.54 & 2786 & Yes \\
 2 & 7.0 & 131072 & 1.3 & 5.67 & 5.98 &   5 & 0.60  & -  & -  &  -   & No  \\
 3 & 7.0 & 131072 & 0.5 & 6.47 & 2.71 &  37 & 0.08  & 3  & 2.55 &  258 & Yes \\ 
 4 & 5.0 & 131072 & 1.3 & 5.20 & 18.36 &  -   &  -  & -  & -  &  -   & No  \\
 5 & 3.0 & 131072 & 1.3 & 4.91 & 39.75 &  -   &  -  & -  & -  &  -   & No  \\
\end{tabular}
\tablecomments{1: The first column indicates the cluster model, followed by 
the dimensionless central potential $W_0$ in the 2nd column. The number of stars in the 
cluster and the half-mass radius are given in the 3rd and 4th columns, respectively. 
The 5th column shows 
 the logarithm of central density followed by the logarithm of the central
 relaxation time. The 7th column gives the total number of collisions that
 occur up to 3 Myrs, followed by the average time between collisions. The 9th
 and the 10th columns indicate the number of collisions leading to runaway
 mergers and the time when runaway merging starts. The mass of the runaway
 star produced at the  end of the runaway merging process is given in the 11th
 column. The last column shows whether runaway
 merging happens or not.}

\end{center}
\end{table}

\onecolumn
\begin{table}
\begin{center}
\caption{Properties of clusters with initial mass segregation introduced within a certain radius \label{tab2}}
\begin{tabular}{cccccccccccc}
\tableline\tableline
 ${}_1$ & ${}_2$ & ${}_3$ & ${}_4$  & ${}_5$  & ${}_6$  & ${}_7$  &
 ${}_8$ &
 ${}_9$ & ${}_{10}$ & ${}_{11}$ & ${}_{12}$ \\
 $Model$ & $W_0$ & $N_{Star}$ & $r_h$ & $M_{IMS}$  & $m_{min}$ & $log~\rho_c$ &
$T_{\rm rel,c}$ &  $Col$ & $\langle T_{\rm col} \rangle$ & $Col_{\rm rm}$ & $RM$ \\
       &       &         & $(pc)$ &$(r \le R_{005})$    & $(M_{\odot})$& $(M_{\odot}/pc^3)$ &
$(Myr)$ &    & $(Myr)$ & & $(Y/N)$ \\ 
\tableline

 6 & 7.0 & 124420 & 1.3 & 0.05 & 30.0 & 5.52 & 3.54 & 3 & 1.00 & - &  No  \\
 7 & 7.0 & 124305 & 1.3 & 0.05 & 50.0 & 5.52 & 3.78 & 7 & 0.43 & - &  No  \\
 8 & 7.0 & 124201 & 1.3 & 0.05 & 90.0 & 5.52 & 3.84 & 2 & 1.50 & - &  No  \\
\end{tabular}
\tablecomments{2: The first and second columns indicate the cluster model and 
the dimensionless central potential $W_0$. The 3rd column shows the number of stars in the 
cluster followed by the half-mass radius in the 4th column. The 5th column gives 
the fraction of total mass of cluster (which is contained within the 5 \% lagrangian 
radius) where the first scenario of IMS is applied. We choose some of these stars 
randomly and assign them with new masses which are larger than the minimum mass 
indicated in the 6th column. The logarithm of central density and the logarithm of the central
 relaxation time are given in the 7th and 8th columns. The 9th and 10th
 columns indicate the total number of collisions that occur up to 3 Myrs and
 the average time between collisions. The 11th columns gives the number of
 collisions leading to runaway mergers. Here we see that none of these
 collisions leads to a runaway merger process. The last column shows whether
 runaway merging happens or not.}

\end{center}
\end{table}


\onecolumn
\begin{table}
\begin{center}
\caption{Properties of simulated clusters with initial mass segregation introduced below a certain energy \label{tab3}}
\begin{tabular}{cccccccccccc}
\tableline\tableline
 ${}_1$ & ${}_2$ & ${}_3$ & ${}_4$  & ${}_5$  & ${}_6$  & ${}_7$  &
 ${}_8$
 & ${}_9$ & ${}_{10}$ & ${}_{11}$ & ${}_{12}$ \\
 $Model$ & $W_0$ & $N_{Star}$ & $r_h$ & $M_{IMS}$ &$m_{min}$ & $log~\rho_c$ &
$T_{\rm rel,c}$ & $Col$ & $\langle T_{\rm col}\rangle$ & $Col_{\rm rm}$ & $RM$ \\
       &    &       &  $(pc)$  &$(\rm{lowest} E_{tot})$ & $(M_{\odot})$& $(M_{\odot}/pc^3)$ &
$(Myr)$     &   &  $(Myr)$ &  & $(Y/N)$ \\ 
\tableline
 9 & 7.0 & 124420 & 1.3 & 0.05 & 30.0 & 5.59 & 8.24  & 2 & 1.50 & - &  No \\
 10 & 7.0 & 124297 & 1.3 & 0.05 & 50.0 & 5.58 & 8.69  & 6 & 0.50 & - & No  \\
 11 & 7.0 & 118805 & 1.3 & 0.10 & 30.0 & 5.54 & 8.07  & 1 & 3.00 & - & No  \\
 12 & 7.0 & 106669 & 1.3 & 0.20 & 30.0 & 5.47 & 6.55  & 2 & 1.50 & - & No  \\
 13 & 7.0 & 106669 & 0.7 & 0.20 & 30.0 & 6.01 & 3.25  & 12 & 0.25 & - & No  \\
 14 & 7.0 & 106669 & 0.6 & 0.20 & 30.0 & 6.21 & 2.57  & 8 & 0.38 & - & No  \\
 15 & 7.0 & 106669 & 0.5 & 0.20 & 30.0 & 6.45 & 1.96  & 20 & 0.15 & - & No  \\
\end{tabular}
\tablecomments{Same as Table~2 except that the 5th column indicates the
  fraction of stars with lowest total energy which were replaced by 
  massive stars. The minimum masses $m_{min}$ of these stars are given in column 6.}

\end{center}
\end{table}

\end{document}